\documentstyle[12pt]{article}
\def\zid{1\kern-0.36em\llap~1}

\newcommand{\beq}{\begin{equation}}
\newcommand{\ber}{\begin{eqnarray}}
\newcommand{\eeq}{\end{equation}}
\newcommand{\eer}{\end{eqnarray}}

\begin{document}

\begin{titlepage}
\rightline{[SUNY BING 8/12/00] } \rightline{ hep-ph/0106137}
\vspace{2mm}
\begin{center}
{\bf ON TESTING FOR NEW COUPLINGS IN TOP QUARK DECAY}\footnote{For
Proceedings of "IVth Rencontres du Vietnam" .}\\ \vspace{2mm}
Charles A. Nelson\footnote{Electronic address: cnelson @
binghamton.edu  } \\ {\it Department of Physics, State University
of New York at Binghamton\\ Binghamton, N.Y. 13902-6016}\\[2mm]
\end{center}


\begin{abstract}
To quantitatively assay future measurements of competing
observables in $t \rightarrow W^+ b $ decay, \newline we consider
the $g_{V-A}$ coupling values of the helicity decay parameters
versus those for \newline `` $(V-A)$ $+$ Single Additional Lorentz
Structures ". There are 2 dynamical phase-type ambiguities $(S+P)$
and $(f_M + f_E)$.  Associated with the latter $(f_M + f_E)$
ambiguity, there are 3 very interesting numerical puzzles at the
mil level. This evidence for the presence of tensorial couplings
in $t \rightarrow W^+ b $ decay is a consequence of the empirical
value of $m_W/m_t $ and the small, but non-zero, ratio $m_b/m_t$.
Measurement of the sign of $ \vert \eta_L \vert = 0.46$(SM) due to
the large interference between the W longitudinal/transverse
amplitudes would exclude such tensorial couplings. Similarly,
sufficiently precise measurements of both $ \eta_L $ and
${\eta_L}^{'} $ could resolve the analogous dynamical ambiguity in
the case of a partially-hidden  $ T$-violation associated with the
additional $f_M + f_E $ coupling.

\end{abstract}

\end{titlepage}

\section{ Introduction}

In physics at the highest available energies, it is always
important to exploit simple reactions and decays so as to search
for new forces, for new dynamics, and for discrete symmetry
violations. Because the t-quark weakly decays before hadronization
effects are significant, and because of the large t-quark mass,
t-quark decay can be an extremely useful tool for such fundamental
searches. Initial tests of the Lorentz structure and of symmetry
properties of $t \rightarrow W^+ b $ decay will be carried out at
the Tevatron[1], but the more precise measurements will be
possible at the CERN LHC [2] and at a NLC [2].

It is important to be able to quantitatively assay future
measurements of competing observables consistent with the standard
model (SM) prediction of only a $g_{V-A}$ coupling and only its
associated discrete symmetry violations. For this purpose, without
consideration of possible explicit $T$-violation, in Ref.[3] plots
were given of the values of the helicity parameters in terms of a
``$(V-A)$ $+$ Additional Lorentz Structure" versus effective-mass
scales for new physics, $\Lambda_i$, associated with each
additional Lorentz structure.  Recently in Ref.[4], to assay
future measurements of helicity parameters in regard to
$T$-violation, the effects of possible explicit $T$-violation were
reported.  In the present formulation, by ``explicit $
T$-violation", we mean an additional complex-coupling, $ g_i / 2
\Lambda_i $ or $ g_i $, associated with a specific single
additional Lorentz structure, $ i = S, P, S \pm P , \ldots $.  In
effective field theory, $\Lambda_i$, is the scale at which new
particle thresholds or new dynamics are expected to occur;
$\Lambda_i$ can also be interpreted as a measure of a top quark
compositeness/condensate scale.  In measurement of some of the
helicity parameters, the LHC should be sensitive to $ \sim 3 $ \%
and the Tevatron in a ``Run 2B" to perhaps the $ \sim 10 $ \%
level (``ideal statistical error levels") [5].

\section{ Testing for the Complete Lorentz Structure in Absence of Explicit $T$-Violation}

A complete measurement of on-shell properties of the $t
\rightarrow W^+ b $ decay mode will have been accomplished when
the 4 moduli are determined and any 3 of the relative phases of
the helicity
amplitudes $%
A(\lambda _{W^{+}},\lambda _b)$.  The helicity parameters appear
directly in various polarization and spin-correlation functions
such as those obtained in Ref.[5].  Since the helicity parameters
appear directly in the various polarization and spin-correlation
functions, it is clearly more model independent to simply measure
them rather than to set limits on an `` ad hoc" set of additional
coupling constants [3].

In the plots in Refs.[3,4], the values of the helicity parameters
are given in terms of a \newline ``$(V-A)$ + Single Additional
Lorentz Structure''. Generically, in the case of no explicit
$T$-violation, we denote these additional couplings by
\begin{equation}
g_{Total} \equiv g_L+g_X \\
\end{equation}
$$ X= \left\{ \begin{array}{ll} X_c =  \; \mbox{chiral} =
\{V+A,S\pm P,f_M\pm f_E\} \\ X_{nc} =  \; \mbox{non-chiral} =
\{V,A,S,P,f_M,f_E\}. \\
\end{array}
\right. $$ For \hskip1em  $t \rightarrow W^+ b$, the most general
Lorentz coupling is $ W_\mu ^{*} J_{\bar b t}^\mu = W_\mu ^{*}\bar
u_{b}\left( p\right) \Gamma ^\mu u_t \left( k\right) $ where $k_t
=q_W +p_b $, and
\begin{eqnarray}
\Gamma _V^\mu =g_V\gamma ^\mu + \frac{f_M}{2\Lambda }\iota \sigma
^{\mu \nu }(k-p)_\nu + \frac{g_{S^{-}}}{2\Lambda }(k-p)^\mu
\nonumber \\ +\frac{g_S}{2\Lambda }(k+p)^\mu
+%
\frac{g_{T^{+}}}{2\Lambda }\iota \sigma ^{\mu \nu }(k+p)_\nu
\end{eqnarray}
\begin{eqnarray}
\Gamma _A^\mu =g_A\gamma ^\mu \gamma _5+ \frac{f_E}{2\Lambda
}\iota \sigma ^{\mu \nu }(k-p)_\nu \gamma _5 +
\frac{g_{P^{-}}}{2\Lambda }(k-p)^\mu \gamma _5  \nonumber \\
+\frac{g_P}{2\Lambda }%
(k+p)^\mu \gamma _5  +\frac{g_{T_5^{+}}}{2\Lambda }\iota \sigma
^{\mu \nu }(k+p)_\nu \gamma _5
\end{eqnarray}
For $g_L = 1$ units with $g_i = 1$, the nominal size of
$\Lambda_i$ is $\frac{m_t}{2} = 88GeV$, see [3].  In the SM, the
EW energy-scale is set from the Higgs-field
vacuum-expectation-value by the parameter $v=\sqrt{-\mu^2 / \vert
\lambda \vert} = \sqrt{2} \langle 0|\phi |0\rangle \sim 246GeV$.
Lorentz equivalence theorems for these couplings are treated in
Ref.[3]. Explicit expressions for the $A(\lambda _{W^{+}},\lambda
_b)$ in the case of these additional Lorentz structures are given
in Ref. [5].

In Table 1 in the top line are the standard model expectations for
the numerical values of the helicity amplitudes $ A\left(
\lambda_{W^{+} } ,\lambda_b \right) $ for $ t\rightarrow W^{+}b $
decay in $ g_L = 1 $ units. The input values are $m_t=175GeV, \;
m_W = 80.35GeV, \; m_b = 4.5GeV$. The $\lambda_b = 1/2$ b-quark
helicity amplitudes would vanish if $m_b$ were zero. For this
reason, if the SM is correct, one expects that the $A(0, -1/2)$
and $A(-1,-1/2)$ moduli and relative phase $\beta_L$ will be the
first quantities to be determined . The $\lambda_b = 1/2$ moduli
are factors of 30 and 100 smaller in the SM. Throughout this
moduli-phase analysis of top decays, intrinsic and relative signs
of the helicity amplitudes are specified in accordance with the
standard Jacob-Wick phase convention.

Versus predictions based on the SM, two dynamical phase-type
ambiguities were found by investigation of the effects of a single
additional ``chiral'' coupling $g_i$ on the three moduli
parameters $\sigma =P(W_L)- P(W_T),\;\xi =P(b_L)-P(b_R),\; $and
$\zeta =\frac 1\Gamma (\Gamma _L^{b_L-b_R}-\Gamma _T^{b_L-b_R})$.
The quantities $$
\begin{array}{c}
P(W_L)= \mbox{Probability} \; W^{+} \; \mbox{ is longitudinally
polarized,} \; \lambda_{W^{+}}=0 \\ \hspace*{-26mm} P(b_L)= \;
\mbox{Probability} \; b \; \mbox{ is left-handed,} \;
\lambda_b=-1/2
\end{array}
$$ In the SM, the final $W$ boson should be $70 \%$ longitudinally
polarized and the b-quark should be almost completely left-handed
polarized. \newline (1) For an additional $S+P$ coupling with
$\Lambda _{S+P}\sim -34.5GeV$ the values of $(\sigma ,\xi ,\zeta
)$ and of the partial width $\Gamma $ are about the same as the SM
prediction. Table 1 shows that this ambiguity occurs because the
sign of the $A_X(0,-\frac 12)$ amplitude for $g_L+g_X$ is opposite
to that of the SM's amplitude. (2) For an additional $f_M+f_E$
coupling with $\Lambda _{f_M+f_E}\sim 53GeV$ the values of
$(\sigma ,\xi ,\zeta )$  are also about the same as the SM
prediction.  In this case, the partial width $\Gamma $ is about
half that of the SM due to destructive interference. (3) From
consideration of Table 1, a third (non-dynamical) phase ambiguity
can be constructed by making an arbitrary sign-flip in the $b_L$
amplitudes, with no corresponding sign changes in the $b_R$
amplitudes. Its exclusion, as well as determination of the 2
remaining independent relative phases necessary for a complete
amplitude measurement will require direct empirical information
about the $b_R$-amplitudes such as from a $\Lambda_b$ polarimetry
measurement [5] of the $b$-polarimetry interference parameters.
Such measurements will be difficult unless certain non-SM
couplings occur: non-chiral couplings like $V$ or $A$, $f_M$ or
$f_E$ (for $\epsilon_+$), $S$ or $P$ (for $\kappa_0$) can produce
large effects[3].  Two dimensional plots of the type $(\epsilon
_{+,-},\eta _L)$ and $(\kappa _{0,1} ,\eta _L)$, and of their
primed counterparts, have the useful property that the unitarity
limit is a circle of radius $0.5$ centered on the origin.

\section{ Remarks on the Dynamical Phase-type Ambiguities}

Due the dominance of the L-handed amplitudes in the SM, the
occurrence of the two dynamical ambiguities [1] displayed in lower
part of Table 1 is not surprising for these 3 chiral combinations
only contribute to the L-handed b-quark amplitudes as $m_b
\rightarrow 0$. Since pairwise the couplings are tensorially
independent, the $g_{V-A} + g_{S+P} $ \& $g_{V-A} + g_{f_M + f_E}
$ mixtures can each be adjusted to reproduce, with opposite sign,
the SM ratio of the two  $( \lambda_W =0, -1 )$ L-handed
amplitudes.

However, in the case of the $f_M+f_E$ phase-type ambiguity, from
Table 1 there are 3 numerical puzzles at the mil level versus the
SM values. In the upper part, the $A_{+} (0,-1/2)$ amplitude for
$g_L +g_{f_M+f_E}$ has about the same value in $g_L = 1 $ units,
as the $A_{SM} (-1,-1/2)$ amplitude in the SM.  As $m_b
\rightarrow 0$, $ \frac{A_{+} (-1,-1/2) } { A_{SM} (0,-1/2) }
\rightarrow
\frac{m_{t}(m_{t}^{2}-m_{W}^{2})}{\sqrt{2}m_{W}(m_{t}^{2}+m_{W}^{2})}
= 1.0038 $. The other numerical puzzle(s) is the occurrence in the
lower part of the Table 1 of the same magnitude of the two
R-handed b-quark amplitudes $A_{New} = A_{g_L =1} / \sqrt \Gamma $
for the SM and for the case of $g_L +g_{f_M+f_E}$. Except for the
differing partial width, by tuning the magnitude of L-handed
amplitude ratio to that of the SM, the R-handed amplitude's moduli
also become about those of the SM. With $\Lambda_{f_M+f_E}$
determined as in Sec. 5, for the four $A_{New}$ amplitudes $ \vert
A_{+} \vert  - \vert A_{SM} \vert \sim (m_b / m_t ) ^2 = 0.0007 $
versus for instance $ \vert A_{SM} (\lambda_W, 1/2) \vert \propto
m_b $. Of course, the row with SM values is from a ``theory"
whereas the row of $g_L +g_{f_M+f_E}$ values is not. Nevertheless,
dynamical SSB and compositeness/condensate considerations do
continue to stimulate interest [5] in additional tensorial
$f_{M}+f_{E}$ couplings. In Table 1, due to the additional $f_M +
f_E$ coupling, the net result is that it is the $\mu =\lambda
_{W^{+} } -\lambda _b = - 1/2 $ helicity amplitudes $A_{New}$
which get an overall sign change.  Fortunately, a sufficiently
precise measurement of the sign of $ \vert \eta_L \vert =
0.46$(SM) due to the large interference between the W
longitudinal/transverse amplitudes can resolve the $V-A$ and $f_M
+ f_E$ lines of this table. Measurement of the sign of the $\eta_L
\equiv \frac 1\Gamma |A(-1,-\frac 12)||A(0,- \frac 12)|\cos \beta
_L $ helicity parameter will determine the sign of $cos \beta_L $
where $ \beta_L $ is the relative phase of the two
$b_L$-amplitudes.

\section{Consequences of ``Explicit $T$-Violation"}

The helicity formalism is based on the assumption of Lorentz
invariance but not on any specific discrete symmetry property of
the fundamental amplitudes, or couplings.  For instance, for $
t\rightarrow W^{+}b $ and $ \bar{t}\rightarrow W^{-}\bar{b} $ in
the case of $ T$-invariance, the respective helicity amplitudes
must be purely real, $ A^{*}\left( \lambda _{W^+},\lambda _b
\right) =A\left( \lambda _{W^+},\lambda _b \right)$ , $
B^{*}\left( \lambda _{W^-},\lambda _{\bar b }\right) =B\left(
\lambda _{W^-},\lambda _{\bar b }\right) $.  Consequently, all of
the primed helicity parameters [5,4] are zero.. $T$-invariance
will be violated if either (i) there is a fundamental violation of
canonical ``time reversal" invariance, or (ii) there are
absorptive final-state interactions. In the SM, there are no such
final-state interactions at the level of sensitivities considered
in the present analysis. To assess future measurements of helicity
parameters in regard to $T$-violation, Ref. [4] gives plots for
the case of a single additional pure-imaginary coupling, $ i g_i /
2 \Lambda_i $ or $ i g_i $, associated with a specific additional
Lorentz structure. (i) An additional $V-A$ type coupling with a
complex phase versus the SM's $g_L$ is equivalent to an additional
overall complex factor in the SM's helicity amplitudes. This will
effect the overall partial width $\Gamma$, but it doesn't effect
the other helicity parameters. For a single additional gauge-type
coupling $V, A, $ or $V+A$, there is not a significant signature
in $ {\eta_L}^{'} $ due to the  $T$-violation ``masking mechanism"
associated with gauge-type couplings [5]. For example: for an
additional pure imaginary $g_R$ coupling plus $g_L$, $
{\eta_L}^{'} \sim m_b/m_t $. So in [4], to test for the presence
of $T$-violation due to additonal gauge type couplings, there are
plots of the b-polarimetry interference parameters
${\epsilon_+}^{'}$ and $ {\kappa_0}^{'}$, and of the partial width
for $ t\rightarrow W^{+}b $ versus pure-imaginary coupling
constant $ i g_i $.  The respective peak magnitudes are $ \sim
0.23, \sim 0.35 $ for the $V+A$ coupling, and are $\sim 0.16, \sim
0.25$ for the $V, A$ couplings.  (ii) Additional $S \pm P, , f_M
\pm f_E , S, P, f_M,$ or $f_E$ couplings can lead to sizable
signatures in the $ \eta _L^{\prime }\equiv \frac 1\Gamma
|A(-1,-\frac 12)||A(0,- \frac 12)|\sin \beta _L $ helicity
parameter for $ \Lambda_i \le \sim 320 GeV$.  There are also
sizable induced effects (factors of $\ge \sim 2$) of such
additional couplings on the partial width for $ t\rightarrow
W^{+}b $.   Ref.[4]  also displays plots of the b-polarimetry
interference parameters ${\epsilon_+}^{'}$ and $ {\kappa_0}^{'}$
versus $ \Lambda_i $ for each of these couplings, except $ f_M +
f_E $  which produces little effect on these 2 parameters.
However, in most cases, such sizable signatures for explicit
$T$-violation due to a single additional coupling can be more
simply excluded by $ 10\% $ precision measurement of the
probabilities $ P(W_L)$ and $ P(b_L)$.  The W-polarimetry
interference parameters $\eta$ and $\omega$ can also be used as
indirect tests, or to exclude such additional couplings.

\section{Tests for ``Partially Hidden $T$-Violation"}

It is possible that $T$-violation exists in the decay helicity
amplitudes, but nevertheless does not significantly show up in the
values of the moduli parameters.  We call this ``partially hidden
$T$-violation" [4]. Based on the notion of a complex effective
mass scale parameter  $ \Lambda_{X} = \vert \Lambda_{X} \vert
\exp{(-i \theta)} $ where $\theta$ varies with the mass scale
$\vert \Lambda_{X} \vert$, we exploit the dynamical phase-type
ambiguities to construct two simple phenomenological models in
which this happens. When $\sin \theta $ $\geq 0$, the imaginary
part of $ \Lambda_{X }$ could be interpreted as crudely describing
a more detailed/realistic dynamics with a mean lifetime scale
$\Gamma_{X} \sim  2 \vert \Lambda_{X} \vert \sin \theta$ of
pair-produced particles at a production threshold $ Re [ 2
\Lambda_{X } ]$.  In the case of the $f_M+f_E$ ambiguity, over the
full $\theta$ range, this construction preserves the magnitudes'
puzzle of Sec. 3. In Ref. [4] are plots of the signatures for a
partially-hidden $T$-violation associated with the $S+P$ and $f_M
+ f_E$ phase-type ambiguities.  Here we will discuss only the
latter case. The additional $f_M+f_E$ coupling $g_{f_M+f_E} / 2
\Lambda_{f_M+f_E} $ now has an effective mass scale parameter $
\Lambda_{f_M+f_E} = \vert \Lambda_{f_M+f_E} \vert \exp{(-i
\theta)} $ in which $\theta$ varies with the mass scale $\vert
\Lambda_{f_M+f_E} \vert$ to maintain SM values in the massless
b-quark limit for the moduli parameters $ P(W_L), P(b_L) ,$ and $
\zeta$.
For $X=f_{M}+f_{E}$, we require $\frac{|A_{X}(-1,-\frac{1}{2})|}{|A_{X}(0,-%
\frac{1}{2})|}=\frac{|A_{L}(-1,-\frac{1}{2})|}{|A_{L}(0,-\frac{1}{2})|}$
so for $m_{b}=0$ the relationship giving $\theta ( \vert \Lambda_{
f_{M}+f_{E} } \vert ) $  is $\cos \theta \simeq
\frac{m_{t}}{4\Lambda }(1+(\frac{m_{W}}{m_{t}})^{2})$ for
$52.9GeV\leq |\Lambda _{f_{M}+f_{E}}|\leq \infty $ which
correspond respectively to  $ 0\leq \theta \leq \pm \frac{\pi
}{2}$. At the maximum of $ {\eta_L}^{'} $, $ \vert
\Lambda_{f_M+f_E} \vert \sim 63 GeV$.   Where $ {\eta_L}^{'} $ has
the maximum deviation, there is a zero in $\eta_L, \eta, \omega$.
As the $\Lambda$ scale increases, the $\sim 2$ destructive
interference effect in the partial width decreases monotonically.
Sufficiently precise measurement of the W-interference parameters
$\eta_L$ and $ {\eta_L}^{'} $ can exclude such partially-hidden
$T$-violation associated with either of the two dynamical
phase-type ambiguities. [ This work was partially supported by
U.S. Dept. of Energy Contract No. DE-FG 02-86ER40291.]

\begin{center}
{\bf Table Captions}
\end{center}

Table 1: For the ambiguous moduli points, numerical values of the
associated helicity amplitudes $ A\left( \lambda_{W^{+} }
,\lambda_b \right) $. The values for the amplitudes are listed
first in $ g_L = 1 $ units, and second as $ A_{new} = A_{g_L = 1}
/ \surd \Gamma $ which removes the effect of the differing partial
width,
$
\Gamma $ for $ t\rightarrow W^{+}b $. [$m_t=175GeV, \; m_W =
80.35GeV, \; m_b = 4.5GeV$ ].

\end{document}